\documentclass[runningheads]{llncs}
\usepackage[T1]{fontenc}
\usepackage{graphicx}
\usepackage{xcolor}
\usepackage{hyperref}
\usepackage{marvosym}
 \usepackage{amsmath}
\usepackage{amssymb}
\usepackage{color}
\begin{document}
\title{Progressive Attention Guidance for Whole Slide Vulvovaginal Candidiasis Screening}
\titlerunning{Progressive Attention Guidance}
%
%

\author{Jiangdong Cai\inst{1} \and Honglin Xiong\inst{1} \and Maosong Cao\inst{1} \and Luyan Liu\inst{1}  \and Lichi Zhang\inst{2} \and Qian Wang\inst{1}\textsuperscript{(\Letter)} }
\authorrunning{J. Cai et al.}

\institute{School of Biomedical Engineering, ShanghaiTech University, Shanghai, China \email{qianwang@shanghaitech.edu.cn} \and  School of Biomedical Engineering, Shanghai Jiao Tong University, Shanghai, China}

\maketitle             
\begin{abstract}

Vulvovaginal candidiasis (VVC) is the most prevalent human candidal infection, estimated to afflict approximately 75$\%$ of all women at least once in their lifetime. It will lead to several symptoms including pruritus, vaginal soreness, and so on. Automatic whole slide image (WSI) classification is highly demanded, for the huge burden of disease control and prevention. However, the WSI-based computer-aided VCC screening method is still vacant due to the scarce labeled data and unique properties of candida. Candida in WSI is challenging to be captured by conventional classification models due to its distinctive elongated shape, the small proportion of their spatial distribution, and the style gap from WSIs. To make the model focus on the candida easier, we propose an attention-guided method, which can obtain a robust diagnosis classification model. Specifically, we first use a pre-trained detection model as prior instruction to initialize the classification model. Then we design a Skip Self-Attention module to refine the attention onto the fined-grained features of candida. Finally, we use a contrastive learning method to alleviate the overfitting caused by the style gap of WSIs and suppress the attention to false positive regions. Our experimental results demonstrate that our framework achieves state-of-the-art performance. Code and example data are available at \url{https://github.com/caijd2000/MICCAI2023-VVC-Screening}.


\end{abstract}

\section{Introduction}
\label{sec:intro}

Vulvovaginal candidiasis (VVC) is a type of fungal infection caused by candida, which results in discomforting symptoms, including itching and burning in the genital area \cite{gonccalves2016vulvovaginal,sobel2007vulvovaginal}.
It is the most prevalent human candidal infection, estimated to afflict approximately 75$\%$ of all women at least once in their lifetime \cite{benedict2019estimation,willems2020vulvovaginal}, resulting in huge consumption of medical resources. 
Currently, thin-layer cytology (TCT) \cite{koss1989papanicolaou} is one of the main tools for screening cervical abnormalities. 
Manual reading upon whole slide image (WSI) of TCT is time-consuming and labor-intensive, which limits the efficiency and scale of disease screening. 
Therefore, automatic computer-aided screening for candida would be a valuable asset, which is low-cost and effective in the fight against infection.

Previous studies for computer-aided VVC diagnosis were mainly based on pap smears rather than WSIs. For example, Momenzadeh et al. \cite{momenzadeh2017automatic} implemented automatic diagnosis based on machine learning. Peng et al. \cite{peng2021efficiently} compared different CNN models on VVC classification. Some works also applied deep learning to classify candida in other body parts \cite{bettauer2022deep,zielinski2020deep}. In recent years, TCT has become mainstream in cervical disease screening compared to pap smear \cite{li2022diagnostic}. Many systems of automatic computer-aided WSI screening have been designed for cytopathology \cite{zhou2021hierarchical,zhang2022whole}, and histopathology \cite{shao2021transmil,zhang2022dtfd}. However, partially due to the limited data and annotation, screening for candidiasis is mostly understudied.

\begin{figure*}[t]
    \centering
    \includegraphics[scale=0.55]{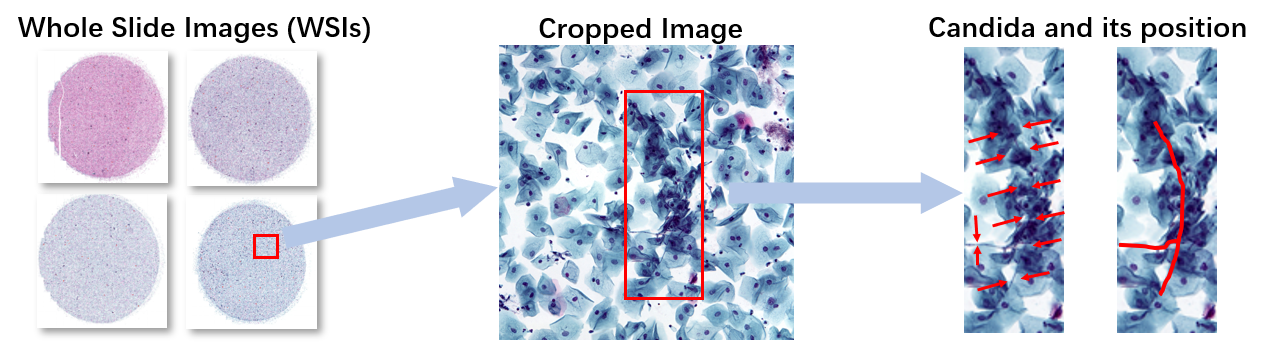}
    \caption{Examples of WSIs (usually about $20000\times20000$ pixels), a cropped image of $1024\times1024$ pixels from the WSI, and zoom-in views of candida and its position (indicated by the red arrows and annotation). }
    \label{fig:Framework}
\end{figure*}

Computer-aided diagnosis for candidiasis through WSI is highly challenging (see examples in Fig.~ \ref{fig:Framework}).
(1) Candida is hard to localize in a large WSI, especially due to its long-stripe shape, low-contrast appearance, and often occlusion with respect to nearby cells. 
The representation of candida is easily dominated by other objects in deep layers of a network. 
(2) In addition to occupying only a small image space for each candida, the overall candida quantity in WSIs is also low compared to the number of other cells. The class imbalance makes it difficult to conduct discriminative learning and to find candida. 
(3) The staining of different samples leads to the huge style gap between WSIs. While collecting more candida data may contribute to a more robust network, such efforts are dwarfed by the inhomogeneity of WSIs, which adds to the risk of overfitting. 
All of the above issues make it difficult for diagnostic models to focus on candida, thus resulting in poor classification performance and generalization capability.

In this paper, we find that the attention for a deep network to focus on candida is the key to the high performance of the screening task.
And we propose a series of strategies to make the model focus on candida progressively. 
Our contributions are summarized into three parts: (1) We use a detection task to pre-train the encoder of the classification model, moving the network's attention away from individual cells and onto candida-like objects; (2) We propose skip self-attention (SSA) to take into account multi-scale semantic and texture features, improving network attention to the candida that is severely occluded or with long hyphae; (3) Contrastive learning \cite{chen2020simple} is applied to alleviate the overfitting risk caused by the style gap and to improve the ability to discern candida.

\section{Method}

\begin{figure*}[t]
    \centering
    \includegraphics[width=1\linewidth]{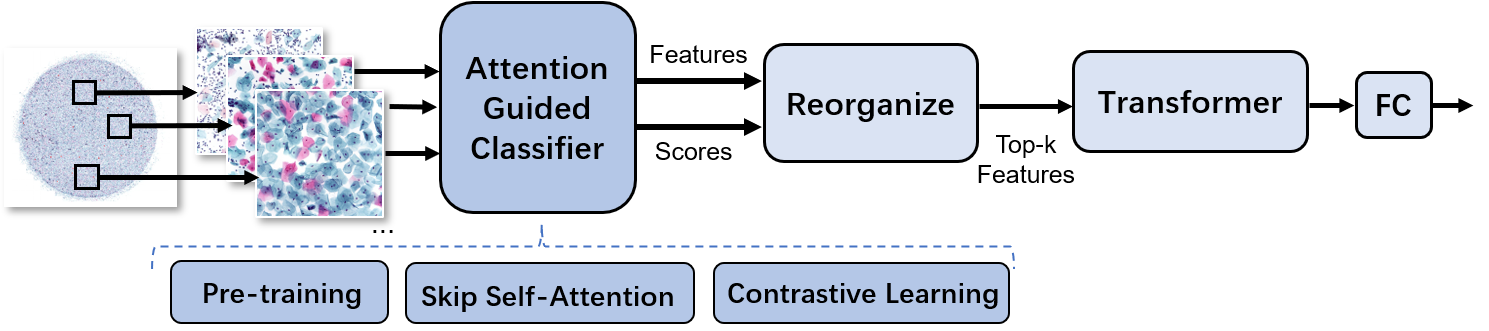}
    \caption{The pipeline of proposed WSI-based VVC screening system.}
    \label{fig:pipeline}
\end{figure*}

We use a hierarchical framework for cervical candida screening, concerning the huge size of WSI and the infeasibility of handling a WSI scan in one shot. The overall pipeline of our framework is presented in Fig.~\ref{fig:pipeline}. Given a WSI, we first crop it into multiple images, each of which is sized 1024$\times$1024. 
For each cropped image, we conduct image-level classification to find out whether it suffers from suspicious candida infection. 
The image-level classifier produces a score and feature representation of the image under consideration. 
Then scores and features from all cropped images are reorganized and aggregated by a transformer for final classification by a fully connected (FC) layer.

\subsection{Detection Task for Pre-Training} 
We use a pre-trained detection model as prior to initialize the classification model. 
In experimental exploration, we find that, if we train the detection network directly, the bounding-box annotation indicates the location of candida and can rapidly establish a rough understanding of the morphology of candida.
However, the positioning task coming with detection lacks enough granularity, resulting in relatively low precision to discern cell edges or folding from candida. 
Meanwhile, directly training a classification model is usually easier to converge. However, in such a task, as candida occupies only a few pixels in an image, it is difficult for the classifier to focus on the target. That, the attention of the classifier may spread across the entire image, leading to overfitted training quickly.

\begin{figure*}[t]
    \centering
    \includegraphics[width=1\linewidth]{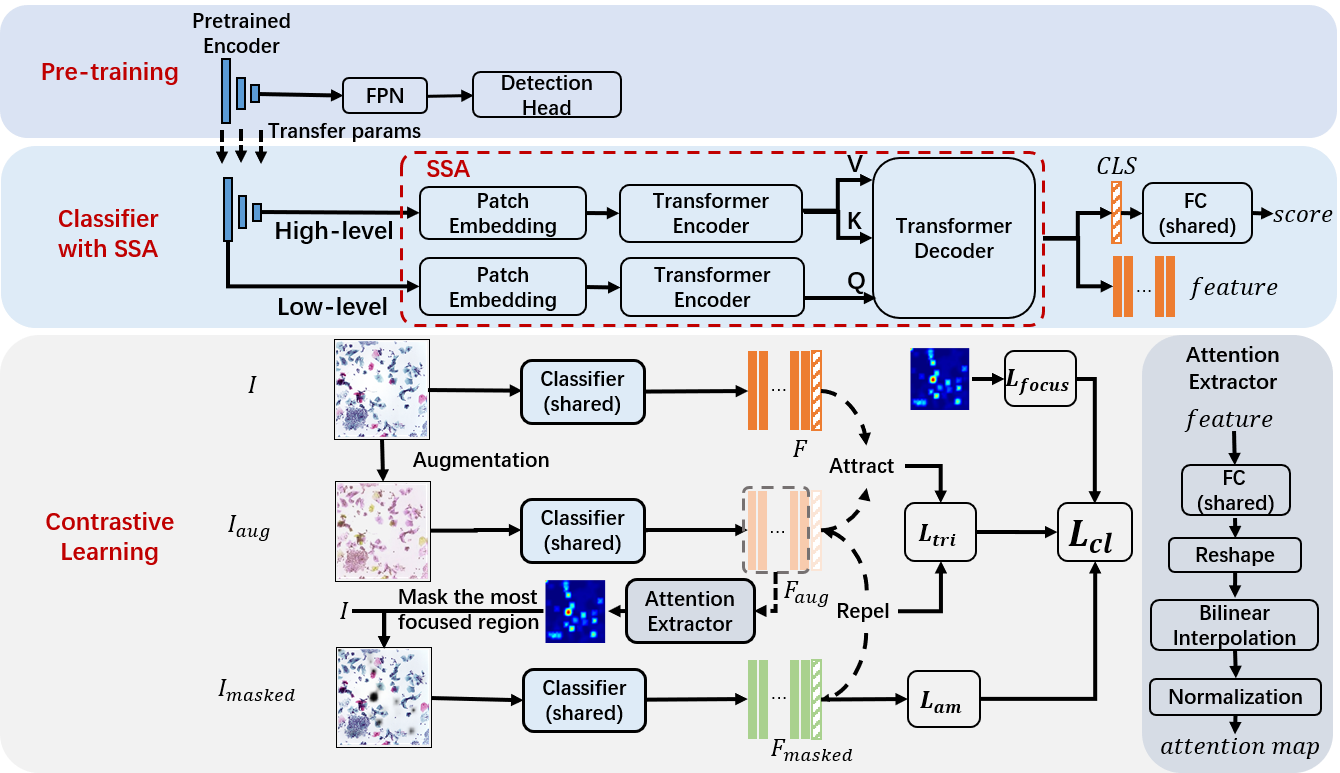}
    \caption{Attention Guided Image-level Classification (corresponding to the classification model in Fig.~\ref{fig:pipeline}). The same parameters are used between modules marked "shared".}
    \label{fig:Training}
\end{figure*}

Therefore, we argue that the detection and classification tasks are complementary to solve our problem. 
Particularly, we pre-train a detector and inherit its advantages in the classifier. 
We use Retinanet\cite{lin2017focal}, which is composed of a backbone attached with FPN (Feature Pyramid Network, FPN, \cite{lin2017feature}) and a detection head, as shown in (Fig.~\ref{fig:Training}). 
We chose the same encoder architecture (Resnet\cite{he2016deep}) for the detection and classification networks. 
To train the encoder with the detection task (Fig.~\ref{fig:Training}), we use bounding-box annotations to supervise Retinanet. 
We then initialize the classification network by directly loading the encoder parameters and freezing the first few layers during the training of the classification network.
Note that pre-training not only discards the complex positioning task but also makes it easier for the classification network to converge especially in the early stage of training.

\subsection{Transformer with Skip Self-Attention (SSA)}  
We design a novel skip self-attention (SSA) module to fuse discriminative features of candida from different scales. At a fine-grained level, the hyphae and spores of candida are usually the basis for judging. Yet we need to distinguish them from easily distorting factors such as contaminants in WSIs and edges of nearby cells. At a coarse-grained level, there is the phenomenon that a candida usually links multiple host cells and yields a string of them. Thus it is necessary to combine long-range visual cues that span several cells to derive the decision related to candida.

CNN-based methods have achieved excellent performance in computer-aided diagnosis including cervical cancer \cite{zhang2022whole}. However, the unique shape and appearance of the candidate incur troubles for CNN-based classifiers, whose spatial field of view can be relatively limited. 
In recent years, vision transformer (ViT) has been widely used in visual tasks for its global attention mechanism \cite{raghu2021vision}, sensitivity to shape information in images \cite{tuli2021convolutional}, and robustness to occlusion \cite{naseer2021intriguing}. 
Nevertheless, such a transformer can be hard to train for our task, due to the large image size, huge network parameters, and huge demand for training data. 
Therefore, to adapt to the shape and appearance of candida, we propose the SSA module and apply it to ViT for efficient learning. 

Specifically, we use the pre-trained CNN-based encoder to extract features for each cropped image. The feature maps extracted after the first layer is considered \textit{low-level}, which contains fine-grained texture information. On the contrary, the feature maps extracted from the last layer are \textit{high-level}, which represents semantics regarding candida. 
To combine the low- and high-level features, we regard the low-level features as queries (Q), and the high-level features as keys (K) and values (V). 
For each patch in the ViT scheme, the transformer decoder computes the attention between low- and high-level features and combines them. The class token 'CLS' is used for the final classification.
The combined feature maps can offer more representative information so that the classifier focuses more on different scales to long-range candida. Meanwhile, the extra SSA structure is simple, which causes a low computation burden.

\subsection{Contrastive Learning}
As mentioned in Section~\ref{sec:intro}, the style gap is another problem, which makes overfitting more severe. In this part, we adopt the strategy of contrastive learning to alleviate such problems and further optimize the attention of the network. Our approach has two key goals: (1) to ensure that the features from the original image remain consistent after undergoing various image augmentations, and (2) to construct an image without the region of candida, resulting in highly dissimilar features compared to the original. 

Inspired by a weakly supervised learning segmentation method\cite{li2018tell}, we construct a contrastive learning method, which will be described in detail in the following sections, as shown in Fig.~\ref{fig:Training}.

To achieve this, we use augmentation and the attention map generated during the training process to construct three types of images and apply contrastive learning to the features extracted from them. 
For a given image $I$, we use image augmentation to generate $I_{aug}$ and use the encoder attached with SSA to extract feature, $F^c_{aug}$. 
The attention map $A$ is transformed from $F^c_{aug}$ by an attention extractor.
 The attention extractor uses FC (the same params as that of the classifier) to reduce the channels of features (except the class token) to 2 (candida and others), then reshape the features representing candida to a feature map, and applies bilinear interpolation to upsample it to the same size of $I$. Eq.~\ref{con:mask} normalizes $A$ to the interval [0,1], obtaining $M$ to represent the likelihood of candida distribution. We get the masked image $I_{masked}$ by subtracting $M$ from $I$. 
\begin{equation}
M= \frac{1}{1+e^{-s(A-\sigma)}} \label{con:mask} \text{,}
\end{equation}
where $\sigma$ and $s$ are used to adjust the range of values, set to 0.5 and 10.

As shown in Fig.~\ref{fig:Training}, the features $F$, $F_{aug}$ and $F_{masked}$ from the three types of images $I$, $I_{aug}$ and $I_{masked}$ by the shared classifier. In our task, we hope that the style gap does not affect the feature extraction of the image, so the distance between $F_{aug}$ and $F$ should be attracted. At the same time, we hope that $F_{masked}$ should not contain the characteristics of candida, which is repelled from $F_{aug}$. To achieve our goal, we introduce triplet loss\cite{schroff2015facenet} for contrastive learning as shown in the first part of Eq.~\ref{con:losscl}.

In addition, we use two constraints, leading to more stable and robust training. 
If our network has effective attention, the masked image should not contain any candida, so the score of the Candida category after the mask $S(I_{masked})$ should be minimized. We use the attention mining loss to handle this, as shown in the second part of Eq.~\ref{con:losscl}. 
Additionally, we need the attention to cover only the partial area around candida, without false positive regions. Otherwise, attention maps that cover the whole image can also result in low $L_{tri}$. We take the average grayscale of attention map $\bar M$ as a restriction, as shown in the last part of Eq.~\ref{con:losscl}.  $L_{tri}$,$L_{am}$, and $L_{focus}$ are combined as $L_{cl}$ to constrain each other and take full advantage of contrastive learning, as shown in Eq.~\ref{con:losscl}.

\begin{equation}
\begin{aligned}
L_{cl}
&= L_{tri}+L_{am}+L_{focus}\\
&=Triplet(F_{aug},F_{orig},F_{masked})+S(I_{masked})+ \bar M \text{.}
\label{con:losscl}
\end{aligned}
\end{equation}

Finally, we use the cross-entropy loss $L_{ce}$ to calculate the classification loss with $labels$. The total loss during training can be expressed as shown in Eq.~\ref{con:lossall}. $\alpha$ is a hyper-parameter, set to 0.1.
\begin{equation}
L =  L_{ce}(S(I_{aug}),labels) + \alpha L_{cl} \text{.}
\label{con:lossall}
\end{equation}

\subsection{Aggregated Classification for WSI} 
With the strategies above, we have built the classifier for all cropped images in a WSI. 
Then we can finish the pipeline of WSI-level classification, which is shown as part of Fig.~\ref{fig:pipeline}.
Specifically, for each cropped image, we conduct image-level classification to find out whether it suffers from suspicious candida infection. 
The image-level classification also produces a score, as well as the feature representation of the image under consideration. 
Then, we reorganize features from all images by ranking their scores and preserving that with top-\textit{k} scores.
We complete the aggregation of the top-\textit{k} features by the transformer and make the WSI-level decision by an FC layer in the final.

\section{Experimental Results}
\subsubsection{Datasets and Experimental Setup.} Our samples were collected by a collaborating clinical institute in 2021. Each sample is scanned into a WSI following standard cytology protocol, which can be further cropped to ~500 images sized 1024$\times$1024. 

For pre-training the detector, we prepare 1467 images with the size of 1024$\times$1024 pixels, all of which have bounding-box annotations. The ratio of training and validation is 4:1.

For training of the image-level classification model, we use 1940 positive images (1467 of which are used in detector pre-training) and 2093 negative images. All images used to pre-train the detector are categorized as training data here. 
The rest 473 images are split in 5-fold cross-validation, from which we collect experimental results and report later. 
The ratio of training, validation, and testing is 3:1:1. 

At the WSI level, we use two datasets. Dataset-Small is balanced with 100 positive WSIs and 100 negative WSIs. We conduct a 5-fold cross-validation, and the ratio of training, validation, and testing is 3:1:1. 
We further validate upon an imbalanced Dataset-Large of 7654 WSIs. 
There are only 140 positive WSIs in this dataset, which is closer to real world. These two WSI-level datasets have no overlay with the data used to train the above detection and classification tasks.

For implementation details, the models are implemented by PyTorch and trained on 4 Nvidia Tesla V100S GPUs. 
All parameters are optimized by Adam for 100 epochs with the initial learning rate of $3\times 10^{-4}$. 
The batch sizes of the detection task, image-level classification, and WSI-level classification are 8, 8, 16, respectively. 
To aggregate WSI classification, we use top-10 cropped images and their features.
We report the performance using five common metrics: area under the receiver operating characteristic curve (AUC), accuracy (ACC), sensitivity (Sen), specificity (Spe), and F1-score.

\subsubsection{Comparisons for Image-Level Classification.} We conduct an ablation study to evaluate the contribution of pre-training (PT), skip self-attention (SSA), and contrastive learning (CL) for the image-level classification, as shown in Table~\ref{table:image}. 
It is observed that with all our proposed components, the network reaches the highest AUC $97.20$, which is 11.49$\%$ higher than the baseline. 
PT shows improvement in all situations, as a reasonable initial focus provides a solid foundation. 
SSA and CL can bring 2.89$\%$ and 6.38$\%$ improvement respectively compared to the method without each of them. It shows that SSA and CL can perform better when the model already has the basic ability to localize candida, i.e., after PT.

\begin{table}[t]\centering
\caption{Image-level classification results and ablation study on the three contributions of our method. (PT: pre-training; SSA: skip self-attention; CL: contrastive loss).}
\begin{tabular}{ccc|ccccc} \hline 
\textbf{PT} & \textbf{SSA} & \textbf{CL} & \textbf{AUC} & \textbf{ACC} & \textbf{Sen} & \textbf{Spe} & \textbf{F1} \\ \hline
 &  &  & 85.71±2.14 & 79.79±2.25 & 83.20±2.35 & 76.55±3.90 & 79.63±1.66 \\ 
 &  & \checkmark &	87.24±2.19 & 85.28±1.45 & 90.88±1.98 & 80.14±3.07 & 85.43±0.90 \\ 
 &  \checkmark &  & 84.97±3.78 &	81.55±2.59 & 87.30±3.05 & 76.27±4.74 & 81.81±2.63\\ 
&  \checkmark & \checkmark & 87.69±3.77 & 86.54±0.98 & 89.59±0.33 & 83.54±1.80 & 86.85±0.89\\ 
\checkmark&   &  & 89.44±3.27 &	89.32±0.96 & 93.41±1.40 & 85.56±1.27 & 89.21±1.32\\ 
\checkmark&   & \checkmark & 94.31±1.75 &	92.54±2.51 & \textbf{93.85±3.21} & 91.22±3.36 & 92.27±2.58\\ 
\checkmark& \checkmark  &  & 90.82±2.57 &	89.84±0.84 & 92.31±1.41 & 87.54±1.49 & 89.56±1.27\\ 
\checkmark& \checkmark  & \checkmark & \textbf{97.20±1.46}  &\textbf{93.89±1.20} & 92.35±1.75 & \textbf{95.22±0.96} & \textbf{93.47±1.18}\\ \hline
\end{tabular}
\label{table:image}
\end{table}

\begin{figure*}[h]
    \centering
    \includegraphics[width=1\linewidth]{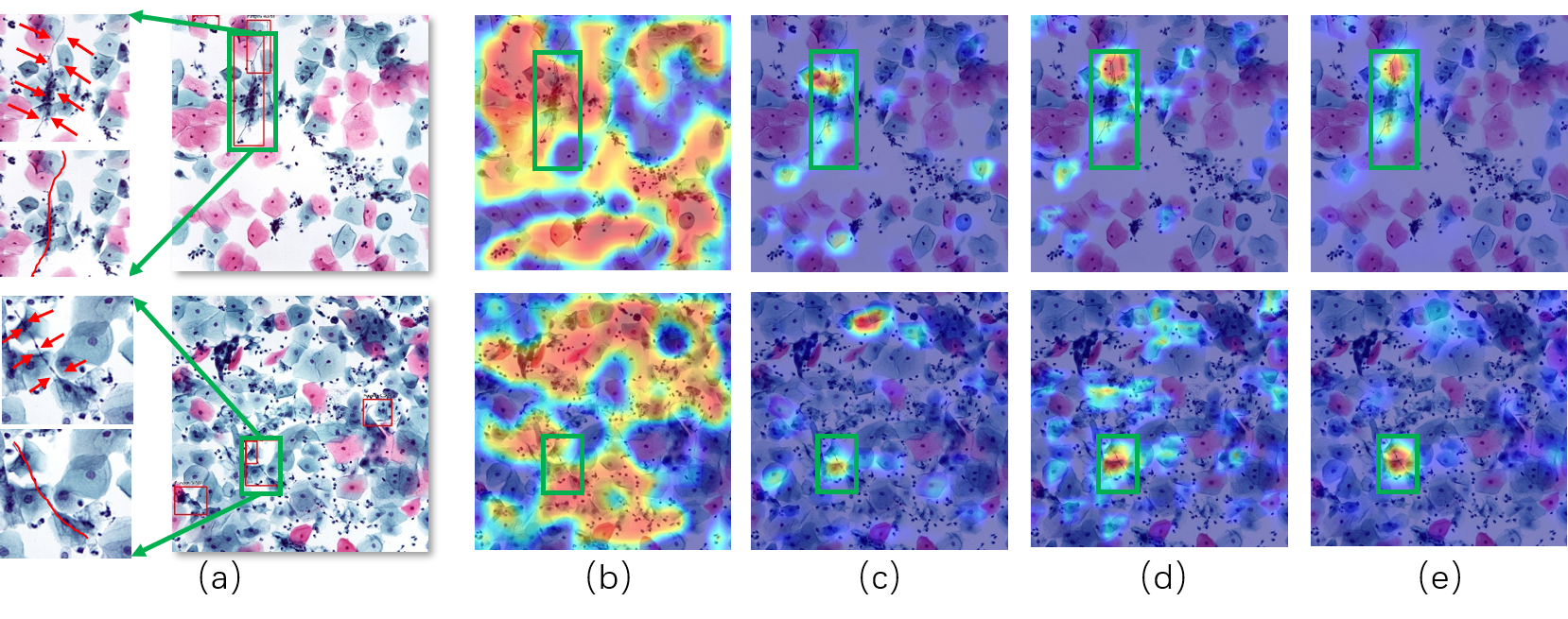}
    \caption{(a) The original image, where the green box indicates candida (enlarged in the left) and the red box shows the prediction of the detection model. Other figures are Grad-CAM of (b) baseline, (c) baseline+PT, (d) baseline+PT+SSA, (e) baseline+PT+SSA+CL.}
    \label{fig:grad-cam}
\end{figure*}

To verify whether our model focuses on important regions of the input image for accurate classification, we visualize the model's attention using Grad-CAM \cite{selvaraju2016grad}. We present two examples in Fig.~\ref{fig:grad-cam}. 
We can see in Fig.~\ref{fig:grad-cam}(b) that the baseline's attention is very scattered spatially. After PT, the model can focus on the candida area, edges of cells, and folds that resemble candida, as shown in Fig.~\ref{fig:grad-cam}(c). After adding the SSA module, more texture information is used to distinguish with cells, as shown in Fig.~\ref{fig:grad-cam}(d). Finally, CL helps the model better narrow its attention, focusing on the most important part as shown in Fig.~\ref{fig:grad-cam}(e). 
These comparisons demonstrate that our proposed method effectively guides and corrects the model's attention. 

\subsubsection{Comparisons for WSI-Level Classification.} We compare our proposed method to other methods in the whole slide of cervical disease screening. To save computation, we did not verify the performance of the methods that performed too poorly on Dataset-Small. The detection-based method \cite{zhou2021hierarchical} uses a detection network to get suspicious candida and classify WSIs with average confidence. Resnet trained without our method is the same as the baseline in Table~\ref{table:image}. At the WSI level, we compare our method with traditional classifiers and a multi-instance learning method TransMIL \cite{shao2021transmil}. We both considered the original TransMIL with pre-trained Resnet-50 and the modified version with our image-level encoder. Table~\ref{table:sample} shows that our method reaches the highest AUC of $95.78\%$ and is the most stable. Our attention-based method brings $6\%$ improvement of accuracy on Data-Small compared to other methods with the same WSI-level method 'Threshold'. Transformer shows a better capacity of feature aggregation than other WSI-level classifiers, raising the AUC on Dataset-Large to $84.18\%$.

\begin{table}[h]
\centering
\caption{Comparision of different methods for WSI classification.}
\begin{tabular}{cc|ccc|cccl}
\hline
\multicolumn{2}{c|}{ Method} &
  \multicolumn{3}{c|}{Dataset-Small} &
  \multicolumn{3}{c}{Dataset-Large} &
   \\ \hline
Image-level &
  WSI-level &
  AUC &
  ACC &
  Sen &
  AUC &
  ACC &
  Sen &
   \\ \hline
Detection &
  Threshold &
  88.57±9.56 & %
  80.00±10.0 & %
  79.03±14.4 &%
  \textbackslash{} &
  \textbackslash{} &
  \textbackslash{} &
   \\
Resnet &
  Threshold &
  88.75±6.58 &%
  77.50±9.35 &%
  82.17±13.0 &%
  \textbackslash{} &
  \textbackslash{} &
  \textbackslash{} &
   \\

Resnet &
  TransMIL &
  93.85±3.71 &%
  89.50±3.67 &%
  \textbf{87.99±7.55} &%
  80.46 &
  86.33 &
  67.85 &
   \\

Ours &
  Threshold &
  92.50±5.36 &%
  86.00±6.44 &%
  83.04±8.71 &%
  82.59 &
  \textbf{88.14} &
  64.29 &
   \\

Ours &
  MLP &
  94.36±3.50 &%
  91.00±6.44 &%
  85.11±13.16 &%
  83.40 &
  87.32 &
  67.86 &
   \\

Ours  &
  TransMIL &
  95.35±1.48 &%
  91.00±3.21 &%
  85.74±4.25 &%
  81.19 &
  86.78 &
  62.86 &
   \\

Ours &
  Transformer &
  \textbf{95.78±2.25} &%
  \textbf{91.64±3.17} &%
  85.55±5.91 &%
  \textbf{84.18} &
  87.67 &
  \textbf{68.57} &
   \\ \hline
\multicolumn{1}{l}{} &
  \multicolumn{1}{l}{} &
  \multicolumn{1}{l}{} &
  \multicolumn{1}{l}{} &
  \multicolumn{1}{l}{} &
  \multicolumn{1}{l}{} &
  \multicolumn{1}{l}{} &
  \multicolumn{1}{l}{} &
   \\
\multicolumn{1}{l}{} &
  \multicolumn{1}{l}{} &
  \multicolumn{1}{l}{} &
  \multicolumn{1}{l}{} &
  \multicolumn{1}{l}{} &
  \multicolumn{1}{l}{} &
  \multicolumn{1}{l}{} &
  \multicolumn{1}{l}{} &
  
\end{tabular}
\label{table:sample}
\end{table}

\section{Conclusion}

We introduced a novel attention-guided method for VVC screening, which can progressively correct the attention of the model. We pre-train a detection task for the initialization, then add SSA to fuse features from coarse and fine-grained, and finally narrower attention with contrastive learning. After obtaining accurate attention and good generalization for the image-level classifier, we reorganized and ensemble features from slices, and make a diagnosis. Both numerical metrics and visualization results show the effectiveness of our model. In the future, we would like to explore the method of weakly supervised learning to make use of a huge number of unlabeled images and jointly train the image-level and WSI-level models.

%
%
%
%
{\small
\bibliographystyle{splncs04}
\bibliography{egbib}
}
\end{document}